\documentclass[aps,pre,twocolumn,groupedaddress,nofootinbib,showpacs,]{revtex4-1}

\usepackage{graphicx}
\usepackage{dcolumn}
\usepackage{bm}
\usepackage{epsfig}
\usepackage{amsmath}
\usepackage{amsfonts}
\usepackage{amssymb}
\usepackage{multirow}
\usepackage{afterpage}
\usepackage{setspace}

\renewcommand{\thetable}{{\bf \arabic{table}}}   
\renewcommand{\thefigure}{{\bf \arabic{figure}}}

\begin{document}
\title{Tuning the Electronic Structure of Anatase Through Fluorination} 

\author{Dario Corradini}
\email{Correspondence and request for materials should be addressed to D. C. (dario.corradini@ens.fr)}
\altaffiliation{Current address: Laboratoire PASTEUR, UMR 8640 ENS--CNRS--UPMC Paris 6, D\'epartement de Chimie, \'Ecole Normale Sup\'erieure, 75005 Paris, France.}
\author{Damien Dambournet}
\author{Mathieu Salanne}
\email{Correspondence and request for materials should be addressed to M. S. (mathieu.salanne@upmc.fr).}
\affiliation{Sorbonne Universit\'es, UPMC Univ Paris 06, CNRS, UMR 8234, PHENIX, Paris, France}


\begin{abstract}
\noindent
A highly fluorinated anatase lattice has been recently reported, providing a new class of materials whose
general chemical formula is $\rm Ti_{1- \mathit x}\square_{\mathit x}X_{4\mathit x}O_{2- 4\mathit x}$ (X$^-$ = F$^-$ or OH$^-$).
To characterise the complex structural features of the material and
the different F environments, we here apply 
a computational screening procedure. After deriving a 
polarisable force--field from DFT simulations, we screen in a step-wise 
fashion a large number of possible configurations differing in the positioning of
the titanium vacancies ($\square$) and of the fluorine atoms. At each step only 10~\% of the configurations  are retained. At the end of the screening procedure, a configuration is selected and simulated using DFT-based molecular dynamics. This allows us to analyse the atomic structure of the material, which is strongly disordered, leading to a strong decrease (by 0.8~eV) of the band gap compared to conventional anatase. 
\end{abstract}

\maketitle 

Titanium dioxide, $\rm TiO_2$, is a widely studied material. $\rm TiO_2$ has in fact several promising applications, for example in the 
fields of photocatalysis, green chemistry and 
energy storage~\cite{chen_titanium_2007,kavan_electrochemical_1996,oregan_a_1991,fujishima_electrochemical_1972,
hoffmann_environmental_1995,kudo_heterogeneous_2008,ravelli_photocatalysis._2009,kamat_tio2_2012}.
Naturally occurring polymorphs of $\rm TiO_2$ include rutile, anatase and brookite. Recently, interest in the polymorphs of
$\rm TiO_2$  has been sparked in particular by their possible application as anodic materials in Li ion batteries~\cite{wagemaker_large_2007,yang_tio2_2012,
morgan_role_2011,yildirim_concentration-dependent_2013,morgan_lithium_2012,morgan_gga+u_2010}.  
Fluorinated $\rm TiO_2$ has also been investigated~\cite{tosoni_theoretical_2013,tosoni_electronic_2012,samadpour_fluorine_2011,czoska_nature_2008,wang_2011,wang_2013} since the presence of F in the compound might improve the sought characteristics of the material~\cite{samadpour_fluorine_2011} or stabilise the highly reactive \{001\} facets of the anatase crystal~\cite{yang_anatase_2008,wang_2011}. The nature of the fluorinated compound depends strongly on the fluorination technique employed~\cite{czoska_nature_2008,yang_fluorine_2010}. So far, the stabilisation of fluorine within the anatase
lattice of $\rm TiO_2$ has been poorly understood, probably because of  the structural complexity of the fluorinated material.

Pure anatase is a tetragonal crystal, with $c\simeq 2.5 a$, and its lattice is characterised by $\rm TiO_6$ octahedral units. 
Recently, a novel synthesis technique conducted in our laboratory~\cite{damien_2013} has led to the preparation of a highly fluorinated anatase material in which fluoride or hydroxide anions replace the oxides in their lattice sites and the resulting charge deficiency is compensated by the formation of a cationic vacancy ($\square$) every four substitutions. The material obtained has thus the general formula $\rm Ti_{1- \mathit x}\square_{\mathit x}X_{4\mathit x}O_{2- 4\mathit x}$, where X$^-$ = F$^-$ or OH$^-$ (the amount of F$^-$ may vary depending on the synthesis conditions). Elemental analysis and synchrotron diffraction have revealed the existence of more than 20\% cation vacancies. In fact the stoichiometric formula 
$\rm Ti_{0.78}\square_{0.22}X_{0.88}O_{1.12}$ has been assigned to the most fluorinated composition of the material.
By using $\rm ^{19}F$ NMR spectroscopy, it has also been possible to discern three different coordination modes for the F atoms: $\rm F-Ti_1\square_2$, $\rm F-Ti_2\square_1$ and $\rm F-Ti_3$, highlighting the complex structural arrangement
present in the material.

Here we report the results of a computational study of the fully-fluorinated, hydroxide-free material (i.e. $\rm Ti_{0.78}\square_{0.22}F_{0.88}O_{1.12}$) performed in order to better
characterise its structural features and the effect of fluorination on the electronic structure. The enormous number of possible structural arrangements of the vacancies and of the F atoms
in the anatase structure render the problem untreatable directly by {\it ab initio} simulations. Therefore we apply a screening
procedure on the possible configurations of the material, in the spirit of the emerging high--throughput techniques~\cite{curtarolo_high-throughput_2013,zakutayev_theoretical_2013,rondinelli_2013}, by using classical Molecular Dynamics (MD).
Several force--fields have been previously proposed for pure $\rm TiO_2$~\cite{catlow_recent_1985,catlow_disorder_1982,mostoller_ionic_1985,
matsui_molecular_1991,sawatari_formation_1982,post_ionic_1986,han_polarizable_2010}.
In this work, we use a {\it polarisable} force--field valid for the pure phase~\cite{corradini_2014_msme} as well as for the fluorinated material. We have extracted its parameters from Density Functional Theory (DFT) simulations, {\it via} a  well-established force and dipole fitting procedure~\cite{salanne_including_2012, tazi_transferable_2012}. We have chosen to derive a new force--field instead of using an already available one for $\rm TiO_2$. This is motivated by the fact that we want the force--field to be compatible with O to F substitutions, as well as with other oxide species, e.g. $\rm SiO_2$, for future studies~\cite{corradini_2014_msme}. The details on the force--field employed are discussed in Supplementary Section S1, while an additional validation of the parameters involving fluorine atoms 
is presented in Supplementary Section S2. 

In order to generate fluorinated samples starting from the pure $\rm TiO_2$ anatase, we apply a screening procedure, similar in spirit to what done by Wilmer {\it et al.} for metal-organic frameworks~\cite{wilmer_large-scale_2012} or by Coudert for zeolites~\cite{coudert2013}.
At the fixed target composition $\rm Ti_{0.78}\square_{0.22}F_{0.88}O_{1.12}$, we consider samples containing F in all possible environments $\rm F-Ti_1\square_2$, $\rm F-Ti_2\square_1$,  and $\rm F-Ti_3$, as suggested by NMR~\cite{damien_2013}. 
We leave the ratio of F in the different environments free to vary at random.
The starting fluorinated structures are generated from the $4\times 4\times 2$ pure anatase 
$\rm TiO_2$ structure~\cite{horn_refinement_1972} ($\rm Ti_{128}O_{256}$) leading to a system thus composed:
$\rm Ti_{100}\square_{28}F_{112}O_{144}$. We generate these configurations by erasing 28 $\rm Ti$ ions at random with no constraints on the creation of adjacent vacancies and  we randomly substitute 112 $\rm O$ with 112 F. We impose that all F and O must be attached to at least one $\rm Ti$. 

The screening procedure is then initiated. The protocol is as follows:

\begin{itemize}

\item[1)] we perform single-point energy calculations on $\simeq 1.5\cdot 10^5$ configurations;
 we then retain the $\simeq 1.5\cdot 10^4$ configurations with the lowest energy for the following step.

\item[2)] we perform 0 K geometry optimisations of the atomic positions, keeping the length of the cell vectors fixed; we retain at maximum the $1.5\cdot 10^3$ configurations with the lowest energy for the following step. 

\item[3)] we perform 0 K cell optimisations of both the atomic positions and the lengths of the cell vectors, while keeping the box angles fixed at  $\alpha=\beta=\gamma=\pi/2$; we retain at maximum the $1.5\cdot 10^2$ configurations with the lowest energy for the following step. 

\item[4)] we temper the configurations performing 10 ps $NVT$ runs at finite temperatures from $T_1=25$~K to $T_{12}=300$~K, every $\Delta T=25$~K. 
 The 15 configurations with the lowest energy at $T_{12}=300$~K are retained for the following step.
\item[5)] for the remaining samples, we perform a series of longer MD simulations at 300~K, first  in the $NVT$ and then in the $NPT$ ensemble.
\item[6)] we then simulate the configuration with the lowest potential energy for 10 ps using using DFT--based molecular dynamics. We extract structural (bond length, fluorine environments) and electronic (density of states) characteristics of the material from this simulation.

\end{itemize}
Testing all the starting configurations in a generic, entirely {\it ab initio} based high-throughput procedure would be impossible. Generally, such studies involve static calculations only since performing {\it ab initio} MD simulations is computationally too expensive. Nevertheless, it is interesting to test whether our selected configurations, i.e., the 10 configurations remaining at the end of step 5) of the screening procedure would have also been selected if {\it ab initio} static calculations had been performed. To test this, we take their 
initial structures and perform a full DFT relaxation. Then we take the same number of random configurations 
from the starting pool of configurations. We find that the configurations given by the classical screening all have a lower
final DFT energy than the ones taken at random. The results of this validation are shown in Supplementary Fig.~S7.

\begin{figure}[htbp]
\includegraphics[clip, scale=0.25]{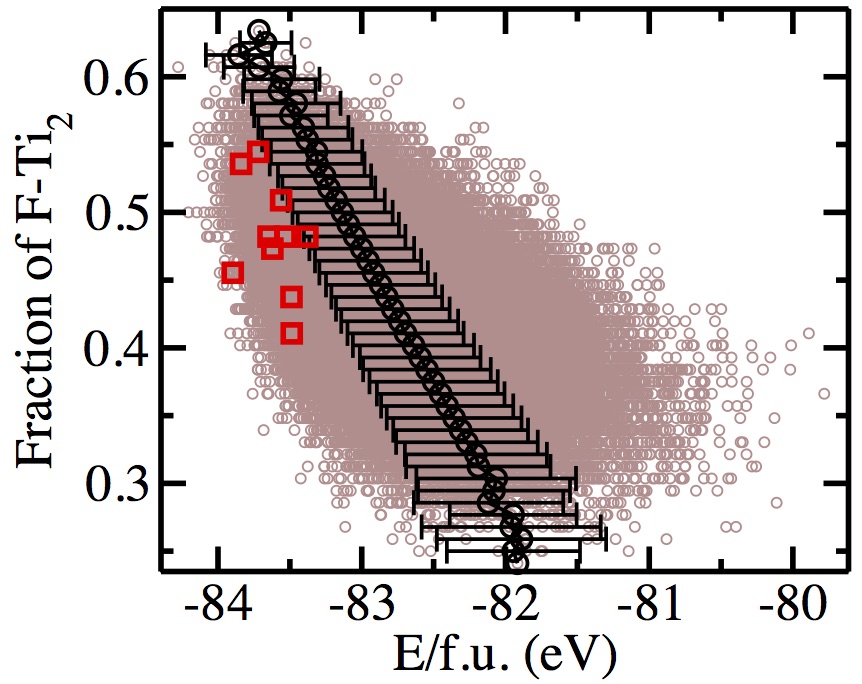}
\caption{{\bf Energy--structure relation.} Fraction of $\rm F-Ti_2\square_1$ vs. energy of the configuration for the initial configurations at 0 K. The points are represented as small brown circles.
For each different value of the fraction of $\rm F-Ti_2\square_1$ we calculate the
mean (black circles) and the standard deviation (black bars) of the corresponding energies. We also report
the values assumed at this stage by the configurations run at the final screening step (red squares).
}
\label{fig:1}
\end{figure}

Next, we analyse how the initial structural arrangements correlate with the energy of the configurations. The results are shown in Fig.~1.
We see that the lowest energies (at 0~K) correlate with a higher fraction of $\rm F-Ti_2\square_1$.
This is consistent with previous static DFT calculations performed on a system with only one vacancy and four O/F substitutions~\cite{damien_2013}, which showed that having the F
closest to the vacancy stabilises the structure.  
In Fig.~1 we also report the initial energies of the best configurations given at the end by the screening
procedure. We observe that these final screening 
configurations are found closer to the average F speciation values rather than at the highest $\rm F-Ti_2\square_1$
relative compositions. Some of them have strongly been stabilised during the procedure, showing the importance of taking into account relaxation and thermal effects (see also Supplementary Fig.~S8).

In order to compare the structural properties of the material as found in the experiments with our simulations, 
we plot together the experimental x-ray structure factor $S(k)$ and the one that we calculate from our trajectories using 
the Ashcroft--Langreth partial structure factors according to the formula:

\begin{equation}
S_{\rm tot}(k)=\frac{\sum_{\alpha\beta} \sqrt{x_\alpha x_\beta}  \, f_\alpha(k) f_\beta(k) S_{\alpha\beta}(k)}{\sum_\alpha x_\alpha f_\alpha^2(k)}
\end{equation}
where $\alpha, \beta= \rm Ti, O, F$. $x_\alpha$ are the relative atomic concentrations of atoms of type $\alpha$, $S_{\alpha\beta} (k)$ are the partial structure
factors calculated from the simulation trajectories using

\begin{equation}
S_{\alpha \beta}(k)=\langle \hat{\rho}_\alpha({\bf k}) \hat{\rho}_\beta^*({\bf k})\rangle
\end{equation}

\noindent where the dynamic variable $\hat{\rho}_\alpha({\bf k})$ represents the Fourier component of the atomic density of type $\alpha$ atoms at wave vector ${\bf k}$:
\begin{equation}
\hat{\rho}_\alpha({\bf k}) = N_\alpha^{-1/2} \sum_{i=1}^{N_\alpha} \exp\left(\imath{\bf k}\cdot{\bf r}_i\right)
\end{equation} 
\noindent with ${\bf r}_i$ the position of atom $i$, and $N_{\alpha}$ the number of atoms of type $\alpha$ in the system. The angular brackets denote a thermal average, which was in practice evaluated as the time average over the whole simulation. Finally, $f_\alpha(k)$ are the $k$-dependent atomic x--ray scattering factors. They are calculated using the analytic approximation:

\begin{equation}
f_\alpha(k)=c_\alpha+\sum_{i=1}^4 a_{\alpha,i} \exp \left[-b_{\alpha,i} \left(\frac{k}{4\pi}\right)^2\right]
\end{equation}
where the coefficients $a_{\alpha,i}$, $b_{\alpha, i}$ and $c_\alpha$ are taken from Ref.~\cite{hemmati_structure_1999} for $\rm O^{2-}$ and from Ref.~\cite{xray_crystall_1992} for $\rm Ti^{4+}$
and $\rm F^-$.  The structure factor calculated from the DFT-based molecular dynamics simulation performed on the final configuration is compared to the experimental signal in Fig.~2. The agreement between the two sets of data is good, taking into account that the experiments have been performed on nanoparticles, which leads to a strong broadening of the peaks, and that part of the fluoride ions are replaced by hydroxide groups. This may also affect the comparison, notwithstanding that F and OH are isoelectronic and thus their contribution to x-ray diffraction should not differ much if they occupy similar sites.

\begin{figure}[htbp]
\includegraphics[clip,scale=0.35]{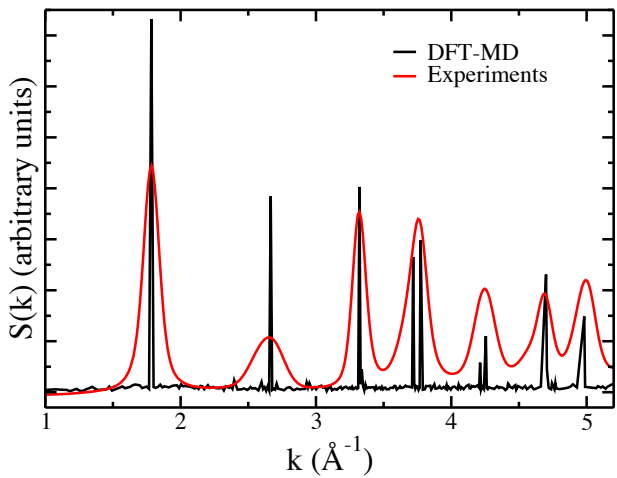}
\caption{{\bf Structure Factor}. Comparison of the structure factor $S(k)$ at ambient conditions measured in experiments (red line) and calculated from a DFT-based molecular dynamics simulation performed on the configuration selected by the screening procedure (black line).
}
\label{fig:2}
\end{figure}

\begin{figure}[htbp]
\includegraphics[clip,scale=0.40]{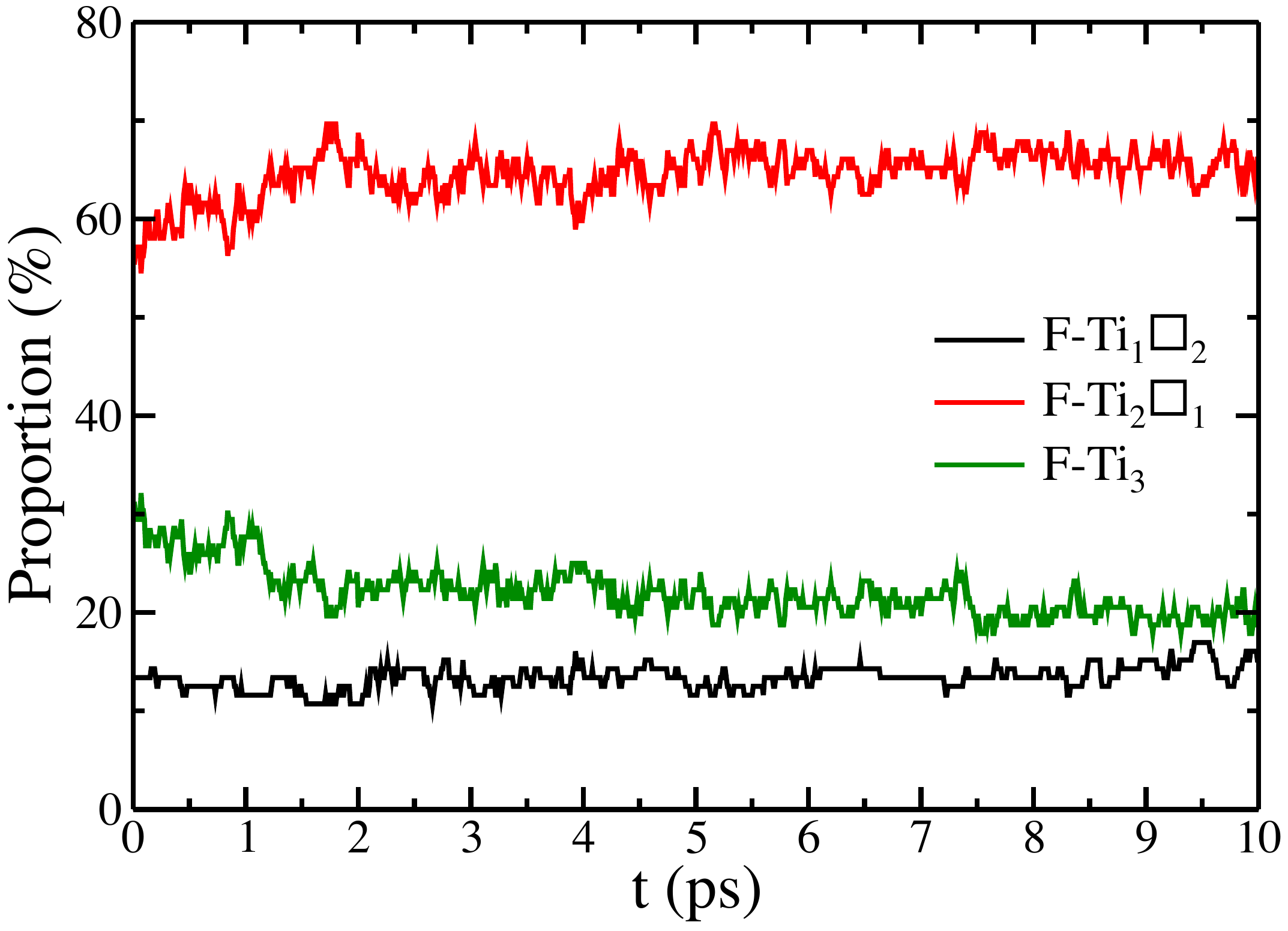}
\caption{{\bf Fluorine coordination number}. Evolution of the proportion of each coordination mode for the fluorine during the DFT--based MD simulation. The cutoff distance for defining Ti--F neighbour atoms is set to 2.7~\AA.
}
\label{fig:3}
\end{figure}

We also calculate the speciation of fluoride during the DFT-based molecular dynamics simulation. Contrarily to the initial configurations analysed in Fig.~1, the local relaxation of the F atoms (especially around the vacancies) leads to a wide distribution of Ti--F distances. It is therefore necessary to introduce a cutoff distance for assigning an environment to the F atoms. In Fig.~3 we show the time evolutions of the concentrations of $\rm F-Ti_1\square_2$, $\rm F-Ti_2\square_1$ and $\rm F-Ti_3$ for a cutoff of 2.7~\AA\, which corresponds to the first minimum of the Ti--F radial distribution function. We observe that after 2~ps of simulation, the concentrations equilibrate around average values of 13/66/21\% for $\rm F-Ti_1\square_2$, $\rm F-Ti_2\square_1$ and $\rm F-Ti_3$ respectively. This compares very well with the percentages measured by NMR in the experimental sample, i.e., 13/70/17\%~\cite{damien_2013}. We can therefore conclude that the structure yielded by our screening procedure is realistic. This allows us to analyse it further in order to predict the material properties. We note that the $\rm F-Ti_2\square_1$ average concentration is larger than the corresponding fraction in the initial pool of configurations as shown in Fig.~1, because the fluorine atoms positions relax around the titanium vacancies during the first 2~ps of the simulation. However, no strong lattice rearrangements are observed, as can be seen from Supplementary Fig.~S9.

\begin{figure}[htbp]
\includegraphics[clip,scale=0.40]{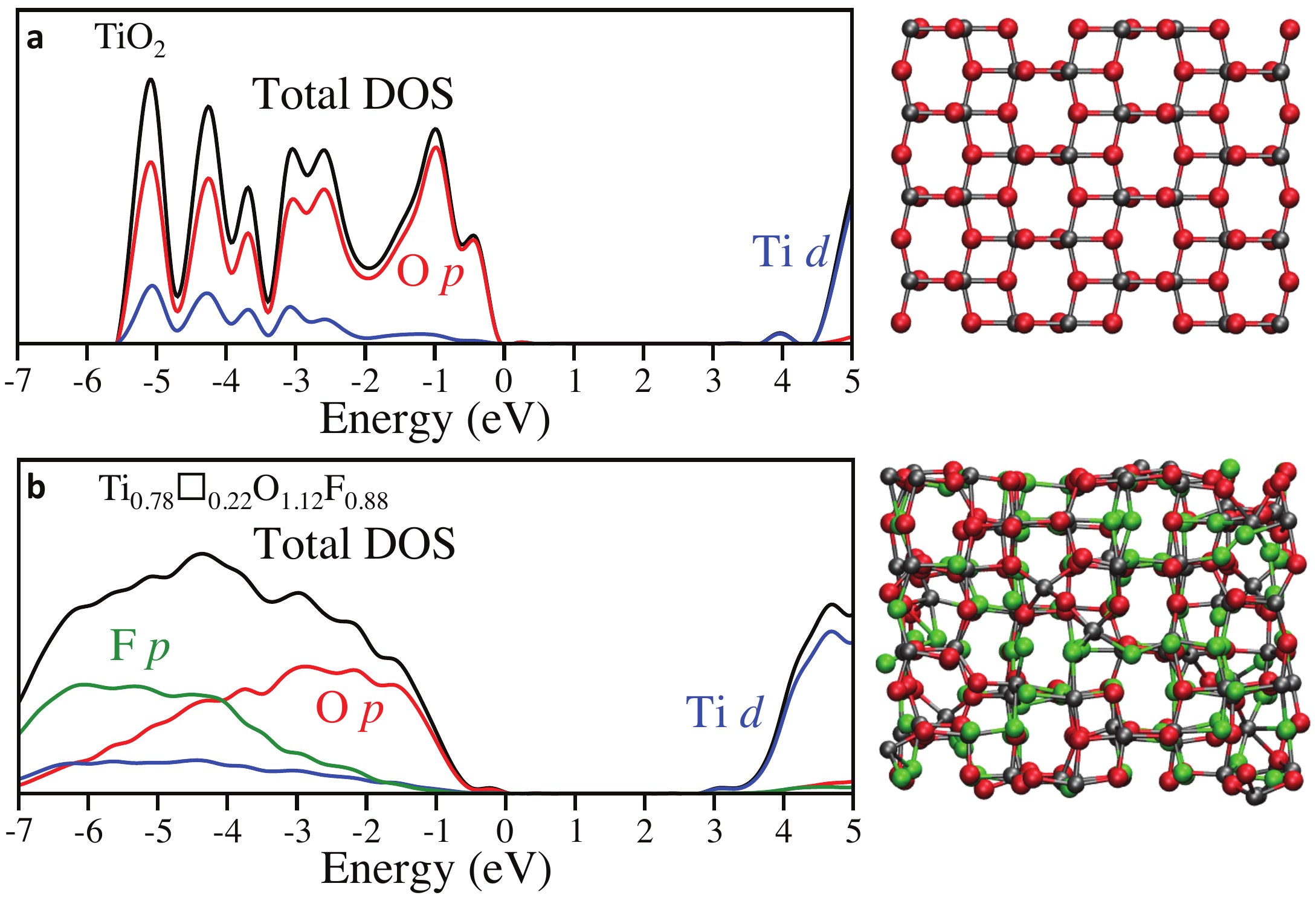}
\caption{{\bf Electronic density of states.} Comparison of the total and ion-decomposed density of states of TiO$_2 $(a) and Ti$_{0.78}$$\square_{0.22}$O$_{1.12}$F$_{0.88}$ (b) calculated using the HSE06 functional. Only the main contributions from the decomposition are shown. The plot for Ti$_{0.78}$$\square_{0.22}$O$_{1.12}$F$_{0.88}$ corresponds to the final snapshot of the simulation, shown on the right side. No significant changes have been observed for other snapshots, see Supplementary Fig.~S10.
}
\label{fig:4}
\end{figure}

The electronic structure is of particular interest for many applications, since TiO$_2$-based materials are widely used in photocatalysis. We have therefore calculated the electronic density of states of fluorinated anatase on a series of snapshots extracted from our DFT-based molecular dynamics simulation, and compared it with the case of pure TiO$_2$ anatase. We have used the hybrid functional HSE06~\cite{hse03,hse06} for these calculations. In agreement with previous works~\cite{scanlon_natmat_2013}, we see in Fig.~4 that the valence band edge of pure TiO$_2$ anatase is dominated by O 2$p$, and the conduction band edge is formed by Ti 3$d$. The band gap is much narrower in $\rm F-Ti_1\square_2$, $\rm F-Ti_2\square_1$, by 0.8~eV. Unlike the case of conventional doping with heteroatoms~\cite{sautet_jpcc_2011}, the additional 2$p$ states associated with fluoride ions do not locate at the top of the valence band, but rather at its bottom. The strong narrowing of the band gap is therefore due to the different structure of the material. In a previous study on TiO$_2$ nanocrystals, Chen {\it et al.} have shown that, due to the presence of structural disorder, their materials exhibit a band gap substantially smaller than the one of pure bulk materials ~\cite{chen_science_2011}. It is very likely that similar effects are at play here.

\begin{figure}[htbp]
\includegraphics[clip,scale=0.55]{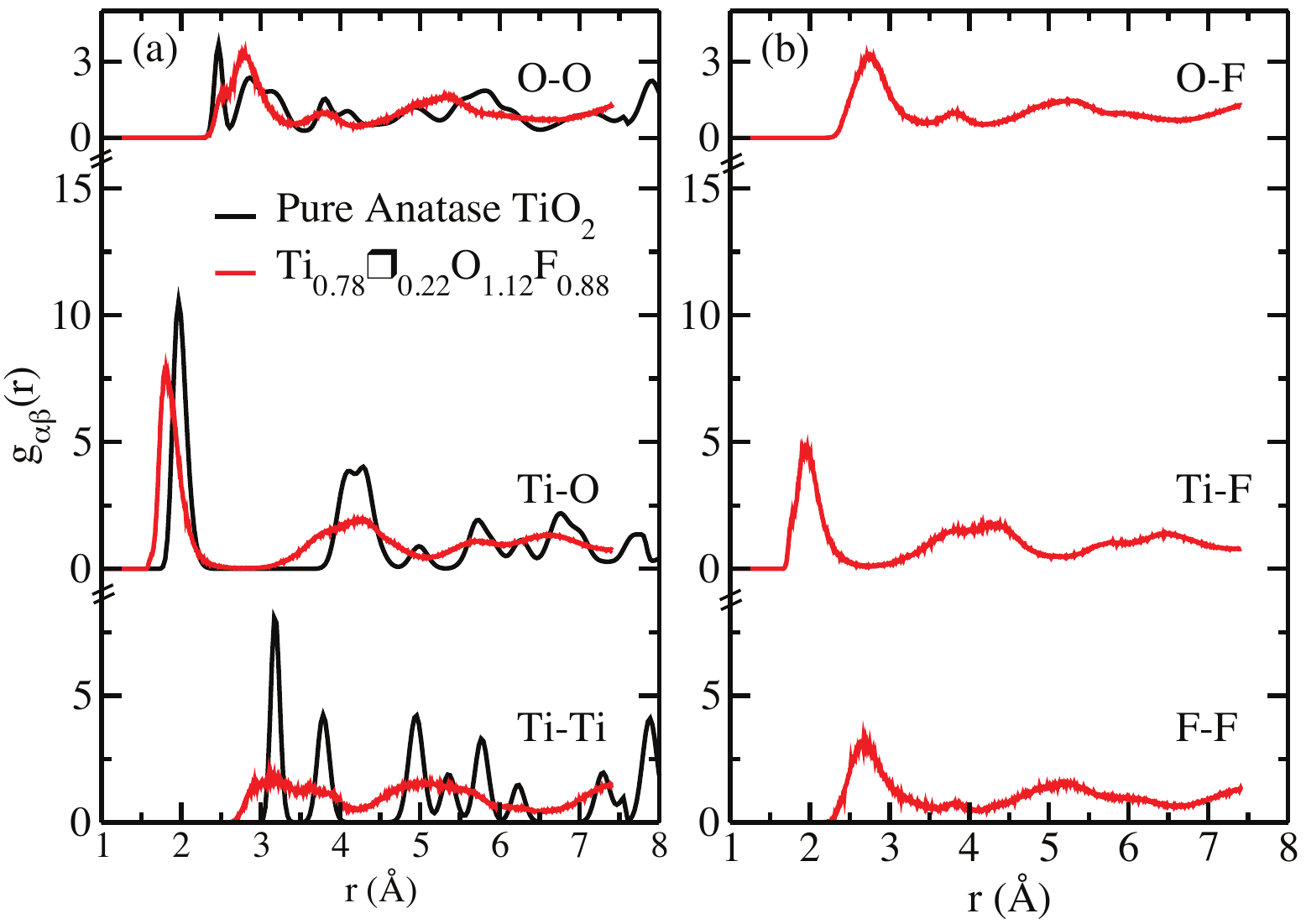}
\caption{{\bf Radial distribution functions.} Partial radial distribution functions $g_{\alpha\beta}(r)$ at ambient conditions calculated from a classical MD simulation of pure 
$\rm TiO_2$ anatase (black) and from a DFT-based simulation of the configuration selected by the screening procedure (red).
}
\label{fig:5}
\end{figure}

The structural disorder is apparently at the origin of these strong electronic structure changes. To test this idea, in Fig.~5 we show the partial radial distribution functions at ambient conditions $g_{\alpha\beta}(r)$ for the simulated pure anatase $\rm TiO_2$ and for the fluorinated anatase configuration selected by the screening procedure. The effect of the disorder introduced by the vacancies is immediately evident looking at the $g_{\rm Ti-Ti}(r)$ and at the $g_{\rm O-O}(r)$.
The $g_{\rm Ti-O}(r)$ structure seems to be conserved to a large extent at least for the first two shells, although the presence of vacancies induces a shortening of the first neighbour distance. The effect of disorder is also clear when looking for example at the region in between the first two peaks.  We also observe that on the one hand, $g_{\rm Ti-O}(r)$ and $g_{\rm Ti-F}(r)$ are very similar, and so are the 
$g_{\rm O-O}(r)$, $g_{\rm F-F}(r)$ and $g_{\rm O-F}(r)$. This confirms that the fluorine atoms substitute the oxygen ones inside the anatase structure.

In conclusion, in order to characterise fluorinated anatase $\rm Ti_{0.78}\square_{0.22}F_{0.88}O_{1.12}$, we have developed a screening procedure employing a polarisable force--field. It has allowed us (i) to select the best configurations starting from a very large pool (hundreds of thousands) of possible configurations; (ii) to reproduce the experimental structure, (iii) to study details of the partial atomic and electronic structure using DFT-based molecular dynamics. Our results show that fluorinated anatase has a highly disordered structure, which results in a lower band gap, by 0.8~eV, compared to conventional anatase. Therefore we conclude that fluorination 
appears as a very promising route for tuning material properties. This may be exploited for several applications, for example photocatalysis. 

\section*{Methods}
We have performed the classical simulations using the software CP2K (single point calculations/geometry/cell optimisations, i.e., steps 1) to 3) of the screening procedure) and the in-house simulation software PIMAIM (molecular dynamics simulations, step 4) and 5) of the screening procedure). We have cut off the short--range interactions at half the norm of the shortest box vector
(or less in $NPT$ runs). The time step for the integration of the equations of motion has been set to 1~fs.

The DFT-based MD simulation has also been performed using the software CP2K~\cite{vandevondele2005a}, using the Quickstep algorithm. We have used the GGA PBE~\cite{perdew1996a} exchange-correlation functional and we have employed the DZVP-MOLOPT-SR-GTH basis set~\cite{vandevondele2007a}. Moreover, we have used the Goedecker-Teter-Hutter~\cite{goedecker1996a} pseudo-potentials; for Ti atoms, the electronic orbitals explicitly represented are 3$s^2$3$p^6$3$d^2$4$s^2$, for O atoms 2$s^2$2$p^4$ and for F atoms 2$s^2$2$p^5$. We have set a plane wave cut-off of 400 Ry. We have added dispersive interactions through the use of the DFT-D3 correction~\cite{grimmed3}, with a cutoff radius of 30~\AA. We have accumulated the trajectory for 10 ps, with the simulations time step being 0.5~fs. We have conducted the simulation in the $NVT$ ensemble with a target temperature of 300~K. We have calculated the electronic density of states on a series of snapshot extracted from the trajectory, using the HSE06 functional~\cite{hse03,hse06}.

\section*{Acknowledgements}

We thank Fran{\c c}ois-Xavier Coudert for introducing us to the screening techniques.
We thank Paul A. Madden, Benjamin J. Morgan and Benjamin Rotenberg for discussions.
The research leading to these results has received funding from the the European Union through the FP7--framework (FLUOSYNES, Contract PCI--GA--2012--321879). We also thank Karena W. Chapman for providing the experimental $S(k)$ data. The work done at the Advanced Photon Source, an Office of Science User Facility operated for the U.S. Department of Energy (DOE) Office of Science by Argonne National Laboratory, was supported by the U.S. DOE under Contract No. DE-AC02-06CH11357.

\section*{Author contributions}
D.C. and M.S. have designed research. D.C. has implemented the screening procedure and has performed most of the simulations. M.S. has conducted the band gap calculations. D.C. and M.S. have written the manuscript and prepared the figures. D.D. has provided the experimental input. All the authors have participated in the discussions and reviewed the manuscript.

\section*{Supplementary information}
In Supplementary Section S1, we describe the analytic form of the classical force--field employed to perform the classical MD simulations
and we tabulate its parameters (see Supplementary Tables~S1 and S2). In Supplementary Section S2, we compare the results reproduced by our classical force--field to DFT--calculated quantities. We first compare the results for the relative stability of structures containing only one vacancy and differing by the positioning of the F atoms (see Supplementary Fig.~S6). Then we compute the DFT--energies (before and after relaxation) of the structures found by our screening procedure and we compare them to the DFT--energies of an equal number of structures selected at random from the pool of starting configurations (see Supplementary Fig.~S7). Supplementary Section 3 deals with the energy distributions at the different screening steps. Fig.~S8 shows the distribution of the energies of the configurations tested and retained at steps 1) to 4) of the screening procedure and Fig.~S9 provides the positions of all the atoms along the DFT--based MD simulation. In Supplementary Section 4, the density of state for 10 different configurations extracted from this trajectory are shown in Fig.~S10. Finally, an example CP2K input is provided in Supplementary Section~S5.

\renewcommand{\thesection}{S\arabic{section}}   
\renewcommand{\thetable}{{\bf S\arabic{table}}}   
\renewcommand{\thefigure}{{\bf S\arabic{figure}}}
\renewcommand{\theequation}{S\arabic{equation}}

\section{Classical model}

We describe the interaction potential between the ions by a classical polarisable force--field whose parameters
we derive from {\it ab initio} DFT simulations. 
Details on how to extract the parameters of the classical force--field from
DFT simulations  are reported in earlier works (see Ref. [34--36] in the main text). In particular, the detailed procedure used for the
pure phases of $\rm TiO_2$ together with its validation are reported in Ref. [34] (of the main text) and are not repeated here.
The force--field parameters for the fluoride ions are obtained in an analogous fashion. We do report here in the following the analytic form of the classical force--field that we use, together with its parameters. 

\subsection{Polarisable ion model (PIM)}

\begin{table*}[!h]
\begin{spacing}{1.5}
\caption{{\bf BMH parameters.} Parameters of the BMH potential extracted from DFT simulations.}
\begin{center}
\begin{ruledtabular}
\begin{tabular}{lccccccc}  Atom Pair & $A_{ij}$ (Ha) & $a_{ij}$ (\AA$^{-1}$)  & $B_{ij}$ (Ha) & $d_{ij} $ (\AA$^{-2}$)  & $C_6^{ij}$ (Ha \AA$^6$) & $C_8^{ij}$ (Ha \AA$^8$)  & $b_D^{ij}$ (\AA$^{-1}$) \\
\hline
O--O  & 290.4 & 4.54668 & -- & -- & 0.48309 & 2.61949 & 2.64562 \\
O--F  & 278.4 & 4.71487 & -- & -- & 0.39890 & 1.55438 & 3.11805\\
O--Ti & 43.0 & 2.86431 & 50,000 & 6.4279 & -- & -- & --\\
F--F  & 282.3 & 4.61849 & -- & -- & 0.32938 & 0.922357 & 3.59048\\
F--Ti &28.3 & 3.13082 & 50,000 & 6.4279 & -- & -- & --\\
Ti--Ti & 1.0 & 9.44863 & -- & -- & -- & -- & --
\end{tabular}
\end{ruledtabular}
\end{center}
\label{tab:1}
\end{spacing}
\end{table*}

\begin{table*}[!h]
\begin{spacing}{1.5}
\caption{{\bf Polarisation parameters.} Parameters of the polarisation part of the interaction potential extracted from DFT simulations.}
\begin{center}
\begin{ruledtabular}
\begin{tabular}{lccc}  Atom / Atom Pair & $\alpha$ (\AA$^{3}$) & $b_{ij} $ (\AA$^{-1}$)  & $c_{ij}$ \\
\hline
O &1.59150\\
O--O & & 4.74888 & 2.227\\
O--F& & -- & --\\
O--Ti & & 3.90122 & 2.13327\\
F & 1.16458\\
F--O & & -- & --\\
F--F & & -- & --\\
F--Ti & & 4.16887 & 2.90678\\
Ti & 0.20442\\
Ti--O & & 3.90122 & -1.90330\\
Ti--F & & 4.16887 & -2.66057\\
Ti--Ti & & -- & --
\end{tabular}
\end{ruledtabular}
\end{center}
\label{tab:2}
\end{spacing}
\end{table*}

The repulsive and dispersive terms of the interactions are taken into account using the 
the  Born--Mayer--Huggins (BMH) form of the interaction potential:

\begin{align}\nonumber
V_{\rm BMHFTD}= &\sum_{i,j>i} A_{ij} e^{-B_{ij}r_{ij}} -f_6^{ij}(r_{ij}) \frac{C_6^{ij}}{r_{ij}^6}\\ &
-f_8^{ij}(r_{ij}) \frac{C_8^{ij}}{r_{ij}^8} \, .
\end{align} 

The damping functions are Tang-Toennies functions of the form
\begin{equation}
f_n^{ij}(r_{ij})=1-e^{b_D^{ij}r_{ij}}\sum_{k=0}^n \frac{(b_D^{ij}r_{ij})^k}{k!}\, .
\end{equation}

When performing molecular dynamics simulations, we add a Gaussian term in the Ti--O and Ti--F interactions that acts as a steep
repulsive wall and accounts for the oxide/fluoride anion hard core:
\begin{equation}
V_{\rm Gaussian} = \sum_{i \in {\rm O,F}, j \in {\rm Ti}} B_{ij} e^{-d_{ij}r_{ij}^2}\, .
\end{equation}
This extra term is used in cases where the ions are
strongly polarised to avoid instability problems at very
small anion--cation separations.

For the Coulombic part of the interaction potential,

\begin{equation}
V_{\rm Coulomb} = \sum_{i,j>i} \frac{q_i q_j}{r_{ij}}\, ,
\end{equation}
the formal charges for the ionic species are used, $-2e$ for O ions, $-e$ for F ions and $+4e$ for Ti ions. 
The many--body electrostatic interactions are described by the induced dipoles $\boldsymbol{\mu}_i$, obtained at each MD step minimising the polarisation energy

\begin{align} \nonumber
V_{\rm pol}= & \sum_i \frac{1}{2\alpha_i} |\boldsymbol{\mu}_i|^2 + \sum_{i,j>i} \left [ \left (q^i \mu_\alpha^j g^{ij}(r_{ij})-q^j\mu^i_\alpha  g^{ji}(r_{ij})\right ) \right. \\
& \left. T_{ij}^\alpha -\mu^i_\alpha \mu^j_\beta T_{ij}^{\alpha\beta} \right ]
\end{align}
where the Einstein summation convention is assumed, $\alpha_i$ is the atomic polarisability and $T$ are the multipole interaction tensors. The damping function
$g_{ij}(r_{ij})$ is of the Tang-Toennies form

\begin{equation}
g_{ij} (r_{ij})=1-c_{ij} e^{-b_{ij} r_{ij}} \sum_{k=0}^4 \frac{(b_{ij}r_{ij})^k}{k!}\, .
\end{equation}

\subsection{Parameterisation}

The repulsion and polarisation parameters of the force--field have been fitted in order to reproduce the forces and dipoles extracted from DFT calculations, using a well-established procedure, see Ref. [34] in the main text. In the present case, we obtain final $\chi^2$ values of 0.16 and 0.37 for the fits of dipoles and forces, respectively. Such values are similar to the ones obtained in our recent work on other oxide materials, i.e. rare earth doped ceria, for example~\cite{burbano2014a}. The dispersion interactions are not taken into account in a proper way in the DFT calculations we have performed. The corresponding parameters have not been fitted, instead they have been taken from our previous works on fluorides and oxides, see Ref. [35] in the main text. Finally, the Gaussian term parameters have been chosen following Marrocchelli {\it et al.}~\cite{marrocchelli2011a}.  The obtained parameters are reported in Table~\ref{tab:1} for the BMH part of the force--field and in Table~\ref{tab:2} for the polarisation part.

\section{Comparison between ab initio and classical simulations}

\begin{figure}[htbp]
\begin{spacing}{1.5}
\includegraphics[scale=0.4,clip]{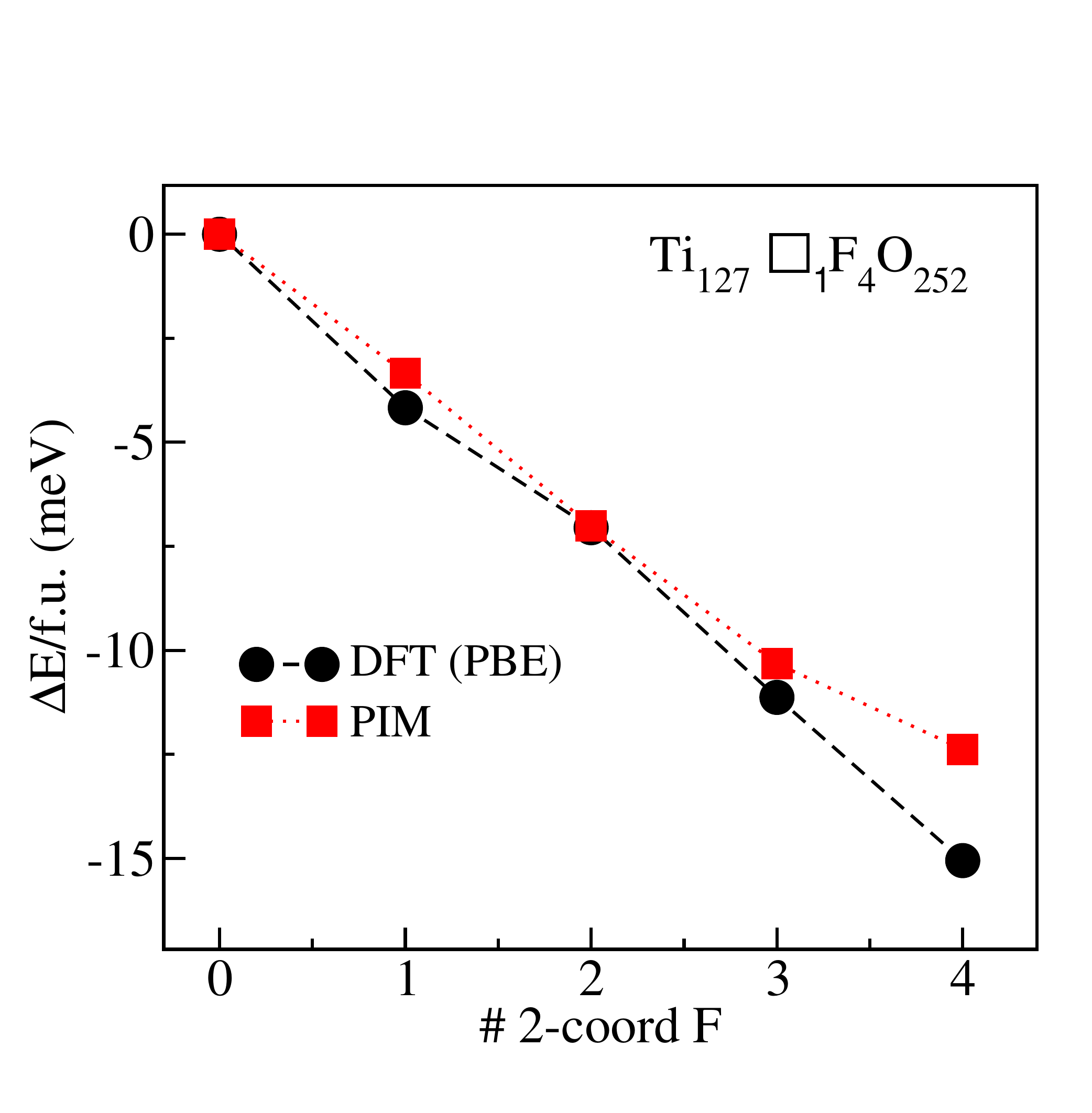}
\caption{{\bf Comparison between DFT and classical potential.} Difference in energy between the case where all F atoms in $\rm Ti_{127}\square_1F_{4}O_{252}$ have
environment $\rm F-Ti_3$ and the cases where F is progressively added in positions neighbouring the vacancies
and has thus $\rm F-Ti_2\square_1$ environment. We have calculated this quantity by DFT (circles) and by our classic
force--field (squares).}
\label{fig:s1}
\end{spacing}
\end{figure}

We assess the behaviour of our force--field in the fluorinated samples, by comparing the results of classical and DFT calculations. For simplicity, we consider the case of one single fluorination, as in Ref. [23] in the main text.
We consider here the system $\rm Ti_{127}\square_1F_{4}O_{252}$ and we distribute the F atoms either at random positions in the lattice (where they substitute O atoms) or at positions neighbouring the cationic vacancy. The latter correspond to 2--coordinated F, i.e. $\rm F-Ti_2\square_1$. We consider the cases where 0, 1, 2, 3 or 4 F are neighbouring the vacancy and have therefore a $\rm F-Ti_2\square_1$ environment, with the remaining F having the 
$\rm F-Ti_3$ environment. We compare the energy calculated in 0 K cell optimisations, using either DFT or our force field. The comparison for the energies is shown in Fig.~\ref{fig:s1}. We see that our classic force-field is able to closely reproduce the decrease in energy with the increase in the number of $\rm F-Ti_2\square_1$, observed by DFT calculations (see also Ref. [23] in the main text). 

\begin{figure}[htbp]
\begin{spacing}{1.5}
\includegraphics[scale=0.35,clip]{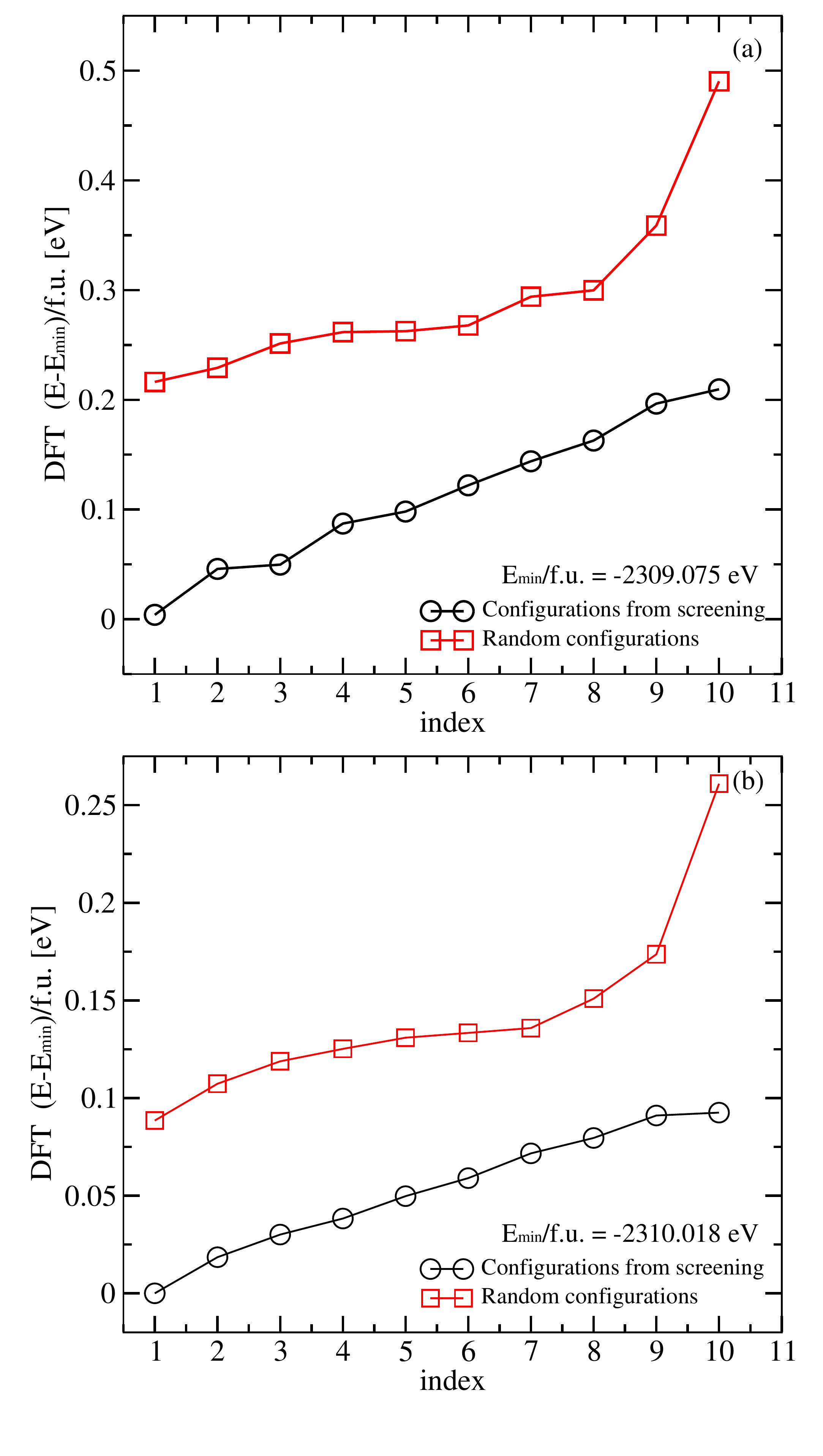}
\caption{{\bf DFT energies of the configurations selected by the screening.} For the configurations left at the end of the screening procedure, we take their initial 
structures and calculate their DFT energy  (black circles) before (a) and after relaxation (b). Those are compared with the DFT energies of the same number of configurations taken at random (red squares) from the initial pools of configuration.
The values of the energies are plotted relative to the lowest DFT energy configuration. Panel~(a) shows the results obtained before the DFT relaxation, panel~(b) shows the results obtained after the DFT relaxation.
}
\label{fig:s2}
\end{spacing}
\end{figure}

Finally in Fig.~\ref{fig:s2} we show the comparison between the  DFT energies of the configurations selected by the screening procedures and the same number of configurations taken at random from the initial pool of configurations of the $\rm Ti_{100}\square_{28}F_{112}O_{144}$, before (panel a) and after (panel b) relaxation of the atomic positions and of the cell dimensions.

\clearpage
\section{SCREENING  PROCEDURE}
\begin{figure}[!h]
\begin{spacing}{1.5}
\includegraphics[scale=0.5,clip]{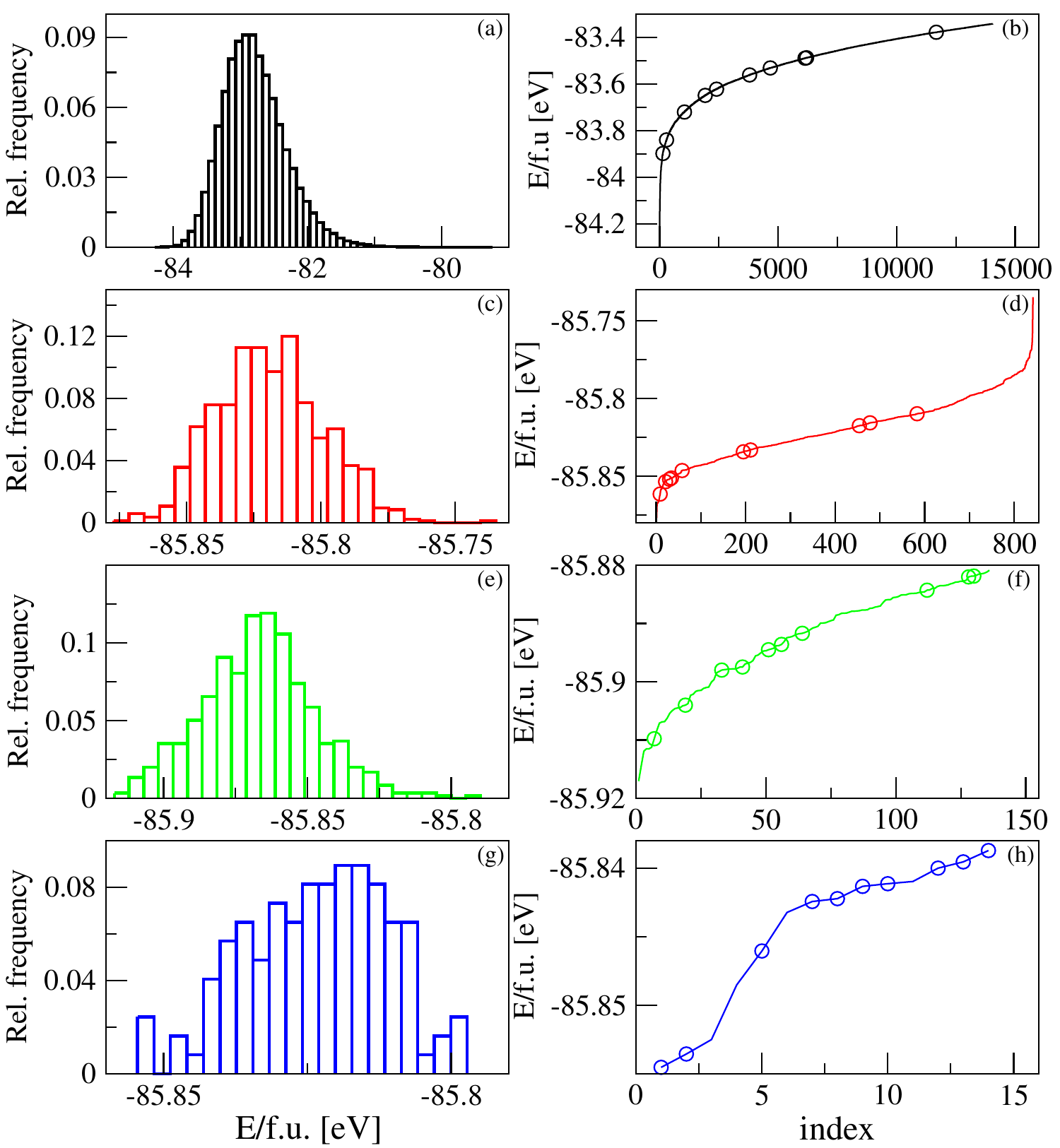}
\caption{{\bf Energy distributions at screening steps 1) to 4).} 
The panels on the left show the relative frequency histograms of the energies calculated for all the configurations tested at each step, while the panels on the right show the sorted energies of the configurations retained after each step. The energies shown are: (a,b) at 0 K for the starting unrelaxed structures; (c,d) at 0 K after the optimization of the atomic positions; (e,f) at 0 K after the optimization of the atomic positions and cell vector lengths; (g,h) at 300 K after the tempering from 25 to 300 K. The open circles in panels (b,d,f,h) indicate
the energy of the configurations left at the end of the screening procedure at each previous step. Note that the number of retained configurations can be less than the target one (see main text). This is due to ``crashed'' unstable configurations that are eliminated from the pool.}
\label{fig:s3}
\end{spacing}
\end{figure}

\begin{figure}[!h]
\begin{spacing}{1.5}
\includegraphics[scale=0.45,clip]{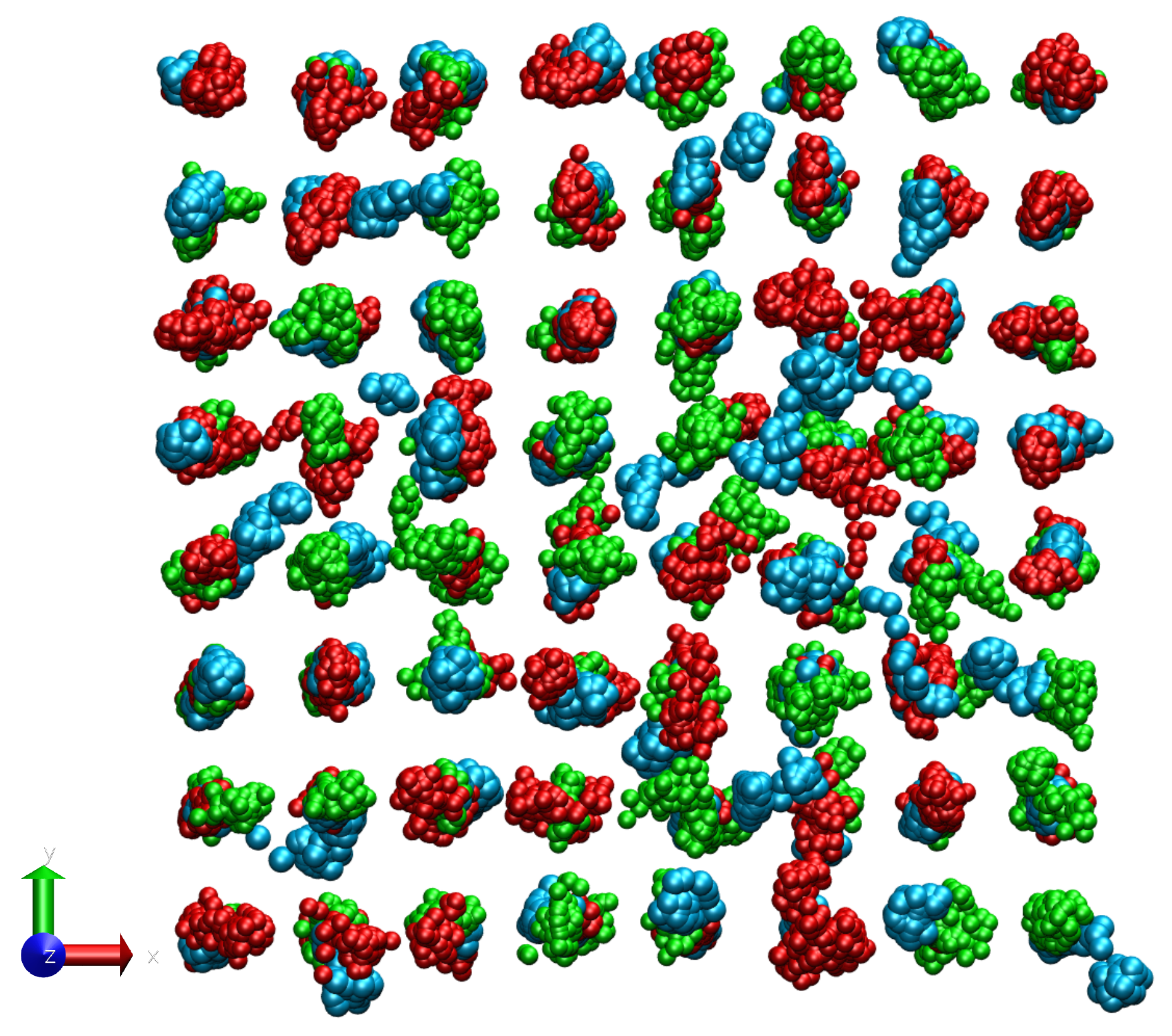}
\caption{{\bf Positions of the atoms along the DFT--based MD simulation.} 
The positions of the Ti (blue), O (red) and F (green) atoms are shown every 10 steps of the simulation. Although structural relaxation close to the vacancies is observed, there is no major lattice rearrangement.}
\label{fig:s4}
\end{spacing}
\end{figure}

\clearpage
\section{Ti$_{0.78}$$\square_{0.22}$O$_{1.12}$F$_{0.88}$ density of states}
\begin{figure}[!h]
\begin{spacing}{1.5}
\includegraphics[scale=0.32,clip]{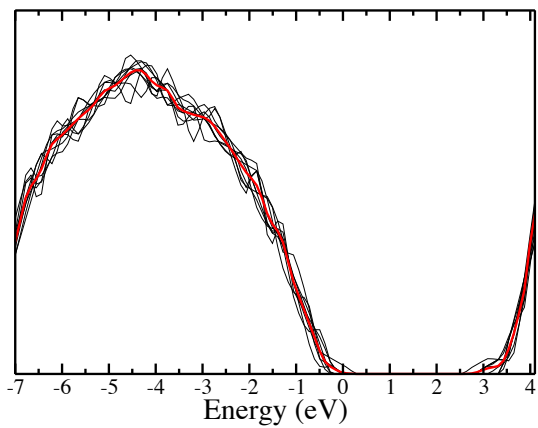}
\caption{{\bf Density of states for several Ti$_{0.78}$$\square_{0.22}$O$_{1.12}$F$_{0.88}$ configurations.} The density of states has been calculated for 10 configurations extracted from the DFT--based molecular dynamics trajectory. The results are shown as thin black lines while their average is shown using a thick red line.}
\label{fig:s5}
\end{spacing}
\end{figure}

\section{CP2K input file}

We here report an example of CP2K input file, prepared for running the system $\rm Ti_{127}\square_1F_{4}O_{252}$ in the $NVT$ ensemble at $T=300$~K, using our classical force--field.

\begin{spacing}{0.8}
\begin{tiny}
\begin{verbatim}

&GLOBAL
 PROJECT 1fn-loc-npt 
 RUN_TYPE MD
 PRINT_LEVEL LOW 
 WALLTIME 50000
 &PRINT
  &EACH 
   MD 500
  &END EACH
 &END PRINT
&END GLOBAL

&MOTION
 &MD
  &THERMOSTAT
   &NOSE
    LENGTH 3
    YOSHIDA 3
    TIMECON 1000.0
    MTS 2
   &END NOSE
  &END THERMOSTAT
 ENSEMBLE NVT
 STEPS 100000
 TIMESTEP 1
 TEMPERATURE 300.0
 &PRINT
  &PROGRAM_RUN_INFO
   &EACH
    MD 500
   &END EACH
  &END PROGRAM_RUN_INFO
  &ENERGY
   &EACH
    MD 500
   &END EACH
  &END ENERGY
 &END PRINT
 &END MD

 &PRINT
  &TRAJECTORY
   &EACH
    MD 5000
   &END EACH
  &END TRAJECTORY
  &RESTART_HISTORY 
   &EACH
    MD 10000
   &END EACH
  &END RESTART_HISTORY
  &RESTART
   BACKUP_COPIES 1
  &END RESTART
  &END PRINT
  
&END MOTION

&FORCE_EVAL
 METHOD FIST
 &MM
  &FORCEFIELD
   &SPLINE
    EMAX_SPLINE 8.0
    RCUT_NB 7.541
   &END SPLINE
  &NONBONDED
   &BMHFTD
    atoms O O
    A 290.4
    B 4.54668
    C 0.483092
    D 2.61949
    BD 2.64562
   &END BMHFTD
   &BMHFTD
    atoms O F
    A 278.4
    B 4.71487
    C 0.3989
    D 1.55438
    BD 3.11805
   &END BMHFTD
   &BMHFTD
    atoms O Ti
    A 43.0004 
    B 2.86431
    C 0.0
    D 0.0
    BD 0.0
   &END BMHFTD
   &BMHFTD
    atoms F F
    A 282.3
    B 4.61849
    C 0.32938
    D 0.922357
    BD 3.59048
   &END BMHFTD
   &BMHFTD
    atoms F Ti
    A 28.3129
    B 3.13082
    C 0.0
    D 0.0
    BD 0.0
   &END BMHFTD
   &BMHFTD
    atoms Ti Ti
    A 1.0
    B 9.44863
    C 0.0
    D 0.0
    BD 0.0
   &END BMHFTD
  &END NONBONDED
  &CHARGE
   atom O 
   CHARGE -2.0000
  &END CHARGE
  &CHARGE
   atom F
   CHARGE -1.000
  &END CHARGE
  &CHARGE
   atom Ti
   CHARGE 4.000
  &END CHARGE
  &DIPOLE
   atom O
   APOL 1.59150
   &DAMPING
    TYPE Tang-Toennies
    ATOM O
    BIJ 4.74888
    ORDER 4
    CIJ 2.227
   &END DAMPING
   &DAMPING
    TYPE Tang-Toennies
    ATOM F
    BIJ 0.0
    ORDER 4
    CIJ 0.0
   &END DAMPING
   &DAMPING
    TYPE Tang-Toennies
    ATOM Ti
    BIJ 3.90122
    ORDER 4
    CIJ 2.13327
   &END DAMPING
  &END DIPOLE
  &DIPOLE
   atom F
   APOL 1.16458
   &DAMPING
    TYPE Tang-Toennies
    ATOM O
    BIJ 0.0
    ORDER 4
    CIJ 0.0
   &END DAMPING
   &DAMPING
    TYPE Tang-Toennies
    ATOM F
    BIJ 0.0
    ORDER 4
    CIJ 0.0
   &END DAMPING
   &DAMPING
    TYPE Tang-Toennies
    ATOM Ti
    BIJ 4.16887
    ORDER 4
    CIJ 2.90678
   &END DAMPING
  &END DIPOLE
  &DIPOLE
   atom Ti
   APOL 0.20442
    &DAMPING
     TYPE Tang-Toennies
     ATOM O
     BIJ 3.90122
     ORDER 4
     CIJ -1.90330
    &END DAMPING
    &DAMPING
     TYPE Tang-Toennies
     ATOM F
     BIJ 4.16887
     ORDER 4
     CIJ -2.66057
    &END DAMPING
    &DAMPING
     TYPE Tang-Toennies
     ATOM Ti
     BIJ 0.0
     ORDER 4
     CIJ 0.0
    &END DAMPING
   &END DIPOLE
  &END FORCEFIELD
  &POISSON
   &EWALD
    EWALD_TYPE EWALD
    EWALD_ACCURACY 1.0e-6
    ALPHA 0.39297 
    RCUT 7.541
    GMAX  19
    O_SPLINE 6
    &MULTIPOLES T
     MAX_MULTIPOLE_EXPANSION DIPOLE
     POL_SCF CONJUGATE_GRADIENT
     EPS_POL 1.0e-6
     MAX_IPOL_ITER 100
    &END MULTIPOLES
   &END EWALD
  &END POISSON
  &PRINT
   &ITER_INFO
    &EACH
     MD 500
    &END EACH
   &END ITER_INFO
  &END PRINT
 &END MM
  &SUBSYS
    &CELL
      ABC 15.082   15.100   20.431
     PERIODIC XYZ 
    &END CELL
    &COORD
 Ti         0.0092526897        2.8105961256        1.2817567910
 Ti         0.0506841050        2.8142375601       11.4965161278
 Ti         0.0373622332        6.5721801714        1.2709321758
 Ti         0.0686762730        6.5665219023       11.4946324231
 Ti         0.0562138844       10.3752296101        1.2665654437
 Ti         0.0930430310       10.3904422990       11.4782100093
 Ti         0.0549241715       14.1400567348        1.2665611962
 Ti         0.0936365683       14.1524822380       11.5041248161
 Ti         3.7759452413        2.7912937424        1.2688107899
 Ti         3.8063797628        2.7956184571       11.4642638411
 Ti         3.8078281440        6.6046747939        1.2696113779
 Ti         3.8492414420        6.6184808818       11.4971070498
 Ti         3.7974199211       10.3708698515        1.2688784727
 Ti         4.0203828276       10.3677305287       11.5495142728
 Ti         3.8655421361       14.1362671796        1.2900093286
 Ti         3.9310964920       14.1262510213       11.5145891174
 Ti         7.5491134334        2.8313499507        1.2719267327
 Ti         7.5681863082        2.8382965342       11.4772300865
 Ti         7.5368925139        6.5978464134        1.2781345496
 Ti         7.5546748184        6.6174658832       11.5111042655
 Ti         7.6040157639       10.3627812967        1.2805033676
 Ti         7.4622388553       10.3668239102       11.5501643194
 Ti         7.6175012880       14.1114312193        1.2650385392
 Ti         7.6209604718       14.1130553468       11.4826702741
 Ti        11.2875516032        2.8265300096        1.2750800406
 Ti        11.3235793462        2.8352807892       11.5014763035
 Ti        11.3484903199        6.5950075403        1.2739718867
 Ti        11.3781032837        6.6029488983       11.4994060714
 Ti        11.3726570777       10.3438744021        1.2627720004
 Ti        11.3869501154       10.3461423554       11.4690935325
 Ti        11.3948366064       14.1469191252        1.2706971376
 Ti        11.4054648836       14.1624354118       11.5083237067
 Ti         1.8955489882        0.9119266960        6.3887525432
 Ti         1.8978237343        0.9133769789       16.5865526041
 Ti         1.9257135506        4.6895911922        6.3785533854
 Ti         1.9404118027        4.7089906429       16.6043602590
 Ti         1.9343509155        8.4920662432        6.3790619262
 Ti         1.9518346955        8.4999711634       16.5905279568
 Ti         1.9525165596       12.2628866292        6.3944507948
 Ti         1.9926244726       12.2501341949       16.5875104054
 Ti         5.6656627583        0.9131930570        6.3865663437
 Ti         5.6901942967        0.9481698804       16.6330334148
 Ti         5.6745443041        4.7141749439        6.3817441594
 Ti         5.6728571349        4.7060333573       16.6151178373
 Ti         5.6902441722        8.4904488086        6.3989824939
 Ti         5.7153330731        8.6552556067       16.5614923347
 Ti         5.7379715177       12.2412941842        6.3973018808
 Ti         5.7423884378       12.0693424778       16.5737394374
 Ti         9.4163802659        0.9350747908        6.3775979904
 Ti         9.4189961791        0.9328323469       16.5879520691
 Ti         9.4321743134        4.7199806153        6.3924056704
 Ti         9.4612587123        4.7409793512       16.5808044328
 Ti         9.4762792944        8.4699692502        6.3887627110
 Ti         9.4708341809        8.4964066142       16.5917323933
 Ti         9.5120319609       12.2363807806        6.3787708348
 Ti         9.5229186076       12.2437664886       16.6039733162
 Ti        13.1755604919        0.9384059638        6.3897255094
 Ti        13.1948430181        0.9502852356       16.5943500922
 Ti        13.2282803150        4.6946721385        6.3859403535
 Ti        13.2413926828        4.7075475043       16.5935058981
 Ti        13.2653352736        8.4621554862        6.3840386973
 Ti        13.2856875386        8.4838322335       16.6183482562
 Ti        13.2613652772       12.2639923032        6.3832808596
 Ti        13.2723944261       12.2572329389       16.6016691772
 Ti         1.8726842508        2.8277145271        3.8381380587
 Ti         1.9277061239        2.8560282353       14.0457425718
 Ti         1.9037779579        6.5985225931        3.8219885884
 Ti         1.9500277893        6.5995357070       14.0377142884
 Ti         1.9505542997       10.3542152293        3.8309291523
 Ti         1.9188210756       10.3586744344       13.9999859258
 Ti         1.9606396537       14.1244640472        3.8294548938
 Ti         2.0081827508       14.1401083529       14.0407335203
 Ti         5.6451516258        2.8211456330        3.8251639215
 Ti         5.6771534617        2.8233289488       14.0137900642
 Ti         5.6939306575        6.5802023077        3.8398661875
 Ti         5.7110707339        6.5163552861       14.0922412885
 Ti         5.7065685497       10.3536592289        3.8328950365
 Ti         5.7259910744       14.1529049523        3.8440195023
 Ti         5.7839305475       14.2542023809       14.1071208952
 Ti         9.4414783466        2.8054100029        3.8378804521
 Ti         9.4601022918        2.8089556560       14.0435886792
 Ti         9.4559310186        6.5808125184        3.8289539910
 Ti         9.4828400504        6.6110676822       14.0401584228
 Ti         9.4686206627       10.3787434428        3.8322819245
 Ti         9.5764227300       10.3957776869       14.0042592707
 Ti         9.4902917028       14.1449715650        3.8183244626
 Ti         9.4994241638       14.1413719375       14.0343092357
 Ti        13.1963868466        2.8028921853        3.8330370903
 Ti        13.2349776080        2.8221974625       14.0457430133
 Ti        13.2086296573        6.6062484093        3.8343715458
 Ti        13.2553567868        6.6389670489       14.0553270443
 Ti        13.2378534778       10.3725492480        3.8225087699
 Ti        13.2727412239       10.3668247532       14.0490364276
 Ti        13.2906480096       14.1262907986        3.8330226768
 Ti        13.3149589939       14.1295796332       14.0592257214
 Ti         0.0066710895        0.9204443731        8.9458796274
 Ti         0.0148939416        0.8884050857       19.1569922011
 Ti         0.0146594993        4.7197127478        8.9348798870
 Ti         0.0132231931        4.6775070873       19.1563708286
 Ti         0.0563794362        8.4793616509        8.9467179417
 Ti         0.0637985346        8.4539855050       19.1521334552
 Ti         0.0916563818       12.2514170022        8.9407169729
 Ti         0.1158856593       12.2076921920       19.1291048606
 Ti         3.7571823911        0.9333130434        8.9444795369
 Ti         3.7386856054        0.8918372625       19.1586663416
 Ti         3.7926965165        4.7202860335        8.9412286575
 Ti         3.7920555593        4.6847533794       19.1618026706
 Ti         3.8300050868        8.4805561564        8.9489178190
 Ti         3.8666398983        8.4446081290       19.1350288637
 Ti         3.8563427476       12.2313301734        8.9646627837
 Ti         3.8768278425       12.2209716162       19.1602455252
 Ti         7.5392813905        0.9386460067        8.9420091757
 Ti         7.5556550309        0.9134362883       19.1636505061
 Ti         7.5790224083        4.6986300311        8.9448911231
 Ti         7.6067245098        4.6715774567       19.1394764767
 Ti         7.6009694465        8.4881799553        8.9637637357
 Ti         7.5831857750        8.4347571603       19.1551848208
 Ti         7.6135486507       12.2488057989        8.9443334140
 Ti         7.5865924905       12.2105094200       19.1472655490
 Ti        11.3219739946        0.9148709362        8.9449524578
 Ti        11.3338759041        0.8891791944       19.1462302817
 Ti        11.3445534617        4.6992552914        8.9490786321
 Ti        11.3455342690        4.6627009070       19.1549642197
 Ti        11.3512777468        8.4906821868        8.9325965141
 Ti        11.3386061373        8.4538641732       19.1394901190
 Ti        11.3794957527       12.2767041122        8.9438094645
 Ti        11.3877843706       12.2362558094       19.1515724146
  O        -0.0022869352        2.6967833227        3.3683648221
  O         0.0529688632        2.7249890910       13.5872398290
  O         0.0422433279        6.3627538345        3.3074707460
  O         0.0739841164        6.3938759503       13.5275939796
  O         0.0365554610       10.4888750817        3.2976670005
  O         0.0701141064       10.6244989097       13.5694870377
  O         0.0571622153       14.3604200491        3.3352859678
  O         0.1043977842       14.2771270428       13.5746828553
  O         3.7873698560        2.5819113948        3.3091705576
  O         3.8091571201        2.5873940265       13.4996670280
  O         3.8535913484        6.8938177594       13.5491807775
  O         3.7995996168       10.5912409401        3.3357654102
  O         3.8495932942       14.0203146696        3.3664805956
  O         3.9398325710       13.9984399455       13.5880148648
  O         7.5257450105        2.9358544584        3.2996974804
  O         7.5714144631        3.0433099259       13.5023921988
  O         7.5446236338        6.8164491877        3.3322298826
  O         7.5521620344        6.8133286564       13.5581050817
  O         7.5914860961       10.2535920603        3.3683201023
  O         7.6326650070       13.9012727510        3.3109861963
  O         7.6237561576       13.8264256202       13.5524590663
  O        11.2926924310        3.0458418563        3.3373461330
  O        11.3280291078        3.0314286238       13.5614436989
  O        11.3359339155        6.4736756110        3.3680000660
  O        11.3784712743        6.6566926809       13.5868645454
  O        11.3810016211       10.1291967496        3.3084725621
  O        11.4172219207       10.0885366363       13.5767292037
  O        11.3750525260       14.2608368682        3.2952312894
  O        11.4035419985       14.2923327473       13.5282467896
  O         1.8920111159        0.7573902469        8.4708043919
  O         1.8922493525        0.6647126339       18.6939360765
  O         1.9249082065        4.5021054792        8.4152607335
  O         1.9352855834        4.4660779020       18.6278358412
  O         1.9296354240        8.6733374780        8.4197987329
  O         1.9510197607       12.4450806147        8.4688323143
  O         1.9853979708       12.4425105425       18.6518300002
  O         5.6687885718        0.7042755875        8.4206872613
  O         5.6874875321        0.7341865795       18.6271776142
  O         5.6721856466        4.8684856876        8.4116553863
  O         5.6933290932        4.5546470328       18.6279847657
  O         5.6882561760        8.6694086480        8.4519524378
  O         5.7130218587        8.7617654367       18.6571104211
  O         5.7420228025       12.1438583761        8.4725575174
  O         5.7395390181       11.9181380627       18.6747228046
  O         9.4165283995        1.0630004009        8.4088330583
  O         9.4386433310        0.7794805591       18.6336511748
  O         9.4357685623        4.9297616176        8.4529935230
  O         9.4606204410        4.9415652150       18.6651012374
  O         9.4820387407        8.3694106539        8.4832545793
  O         9.4699704559        8.2631871171       18.6683678861
  O         9.5175864541       12.0075432038        8.4275568685
  O         9.5178748023       12.0639729860       18.6205482210
  O        13.1778240075        1.1409470039        8.4561034154
  O        13.1922243494        1.1189141237       18.6616416377
  O        13.2303598457        4.5620960393        8.4738481055
  O        13.2389368515        4.5015656723       18.6823813572
  O        13.2675482430        8.2774671659        8.4251400025
  O        13.2832231126        8.2435251580       18.6278012167
  O        13.2590673160       12.3858635902        8.4136473865
  O        13.2886850751       12.1549471037       18.6276477443
  O         1.6792808545        2.8313579725        5.8897982264
  O         1.7309968446        2.8588078968       16.0889854042
  O         1.8568224669        6.5964538101        5.8543641611
  O         1.8003852451        6.6165514370       16.0573277201
  O         2.1791096791       10.3480364548        5.8618907037
  O         2.1855366649       10.3603530656       16.1014918145
  O         2.0840118270       14.1279508357        5.9279629747
  O         1.8851929926       14.1341426634       16.1084858859
  O         5.5928936186        2.8218726549        5.8570279479
  O         5.5340524650        2.8357268602       16.0644138647
  O         5.9359836220        6.5740143227        5.8613706564
  O         5.9014987863        6.5289491195       16.0566550393
  O         5.8202681560       10.3558548891        5.9307551705
  O         5.7395607699       10.3694001153       16.0540031146
  O         5.5223635933       14.1544765572        5.8948778047
  O         5.6231125457       14.2494149144       16.0644610508
  O         9.6738992345        2.7968539690        5.8597515173
  O         9.6647914904        2.8104661575       16.0736590811
  O         9.5683812798        6.5847536109        5.9272009788
  O         9.6560883606        6.6269948919       16.1073127626
  O         9.2775316233       10.3808227189        5.8909912300
  O         9.3542821481       10.4022764574       16.1141004770
  O         9.4255389304       14.1458365807        5.8534221408
  O         9.3869736546       14.1542683026       16.0655712802
  O        13.3044521226        2.8064294914        5.9205765521
  O        13.3154024446        2.8327434087       16.1145121399
  O        13.0040279543        6.6086201058        5.8928910397
  O        13.0450667595        6.6316639614       16.1096598800
  O        13.1935919063       10.3724374564        5.8632602830
  O        13.5327422232       14.1203244950        5.8617411981
  O        13.5404510326       14.1403695445       16.1015207395
  O         0.0612471592        0.9088343860        0.8050433067
  O         0.1179952991        0.9213077396       11.0392238481
  O        -0.2287904771        4.7083932583        0.7871239191
  O        -0.2033464162        4.7212868415       10.9895752919
  O        -0.1162569310        8.4873793996        0.7404316193
  O         0.0857049761        8.4788041005       10.9785000249
  O         0.2576586752       12.2372615705        0.7511297707
  O         3.5038126666        0.9221370013        0.7898121285
  O         3.5459724319        0.9467618494       10.9782941314
  O         3.6138001775        4.7181584512        0.7405744675
  O         3.7912290205        4.7045336809       10.9524248071
  O         4.0034592404        8.4687117719        0.7523578725
  O         4.0710259509        8.4581308931       10.9991293777
  O         3.9296590838       12.2392499993        0.8017644026
  O         3.9725554175       12.2441637859       11.0937779572
  O         7.3912031098        0.9380251398        0.7416202509
  O         7.5782329020        0.9409593883       10.9499206183
  O         7.7435087642        4.7021350515        0.7583014693
  O         7.8163881164        4.6837477924       10.9548317485
  O         7.6293319436        8.4618365414        0.7979761483
  O         7.7190030393        8.4839371711       11.0855558375
  O         7.3477613461       12.2500754723        0.7845011408
  O         7.4061904199       12.2647493439       11.0241208925
  O        11.4730559807        0.9189041921        0.7590220624
  O        11.5608841968        0.9063871791       10.9723721028
  O        11.4119821011        4.6922973540        0.8010296263
  O        11.4668237730        4.7051515940       11.0411284425
  O        11.1038795762        8.4861548884        0.7808478547
  O        11.1316764255        8.4880955344       10.9937760214
  O        11.2120786149       12.2594068045        0.7391778996
  O        11.3844318714       12.2670116707       10.9770301498
  O         1.8638582388        3.0207547943        1.8050674229
  O         1.9010365279        3.0698158565       11.9929414075
  O         1.9021388031        6.7235187383        1.7275217999
  O         1.9635541048        6.5578129882       11.9821581615
  O         1.9428077732       10.1259712645        1.7715154664
  O         1.9669038199       10.1853083303       12.0344792446
  O         1.9621625649       14.0230583287        1.7991284932
  O         2.0189117945       14.1061649542       12.0391725703
  O         5.6430930598        2.9368537844        1.7215709314
  O         5.6870527180        2.8314538579       11.9336483065
  O         5.6851449437        6.3426539268        1.7732417658
  O         5.7129075600        6.3080940781       11.9825848845
  O         5.7041827906       10.2586187515        1.8004393975
  O         5.7465181589       10.2958040985       12.0423500082
  O         5.7161037780       14.3639895941        1.8068634268
  O         5.7724607436       14.4558033101       11.9852306525
  O         9.4332043265        2.5898290849        1.7752760936
  O         9.4778611738        2.5835242806       11.9899370039
  O         9.4512158144        6.4986543822        1.7962524164
  O         9.4835479522        6.4519581189       12.0375689511
  O         9.4569113089       10.5708993466        1.8051843773
  O         9.5304487961       10.5797895198       12.0429086017
  O         9.4867887531       14.2501072868        1.7229812289
  O         9.5052332402       14.2103288138       11.9852400998
  O        13.1939497223        2.7063452250        1.7978635099
  O        13.2385900913        2.7214324277       12.0263046857
  O        13.2001630346        6.8071779285        1.8040399520
  O        13.2490668278        6.8144202341       11.9975439615
  O        13.2357729435       10.4932981426        1.7239104605
  O        13.2789225949       10.3810186722       11.9833955591
  O        13.2817190643       13.8962747069        1.7704703815
  O        13.3185803989       13.9357820521       11.9997203983
  O         0.0023405852        0.8581098685        6.9160387112
  O         0.0240579712        0.9804197871       17.1264332075
  O         0.0073478564        4.9309257918        6.8979744431
  O         0.0191832455        4.9356287295       17.1215147197
  O         0.0524385114        8.5396942664        6.8526404713
  O         0.0711555613        8.6813144991       17.0861304843
  O         0.0927874617       12.0275570696        6.8926365997
  O         0.1152465606       12.0482733232       17.0720920509
  O         3.7530516096        1.1539528285        6.9005092670
  O         3.7430638043        1.1324672046       17.1355796891
  O         3.7929050168        4.7963548776        6.8443887482
  O         3.7977847291        4.9205939605       17.0907612704
  O         3.8284285266        8.2449403208        6.8865815510
  O         3.8739800521        8.3109047342       17.0419938749
  O         3.8490558712       12.1533916216        6.9238322625
  O         3.8899655575       12.3578151791       17.0930152252
  O         7.5352024976        1.0207445705        6.8462281539
  O         7.5627651276        1.1661932778       17.0883930382
  O         7.5796313527        4.4701244005        6.8816455903
  O         7.6080529942        4.5439394644       17.0783036239
  O         7.5991098634        8.4032184390        6.9208535462
  O         7.5923489915        8.4129783570       17.0883755764
  O         7.6081186258       12.4656181289        6.9044719295
  O         7.5842099803       12.4291414330       17.0965718833
  O        11.3240336491        0.6874900449        6.8811611803
  O        11.3338246086        0.7696910132       17.0674170087
  O        11.3406651648        4.6148385577        6.9170639229
  O        11.3573660406        4.6330058237       17.1178247849
  O        11.3469552369        8.7013974844        6.9067341178
  O        11.3517072387        8.7086211279       17.1315670131
  O        11.3785281080       12.3701449175        6.8513142558
  O        11.3956047390       12.4585169455       17.0805807392
  O         0.1434580193        2.8025228952        9.4640377960
  O         0.1798758842        2.7683636351       19.6894656064
  O         0.2565206371        6.5796710886        9.4208163149
  O         0.2494094097        6.5279760806       19.6587878817
  O        -0.1206131415       10.3804461429        9.4340652625
  O         0.1370774878       10.3271503935       19.6114941580
  O        -0.1365454287       14.1590236024        9.4680998144
  O        -0.1405125437       14.1082682127       19.6464458403
  O         3.9795163770        2.8006403770        9.4011406766
  O         3.9530496739        2.7479319430       19.6647668379
  O         3.6278635521        6.6180315224        9.4205598012
  O         3.9227780074        6.5588333695       19.6112743669
  O         3.6903855129       10.3861021241        9.4997885118
  O         3.5690634336       10.3410430819       19.6519069714
  O         3.9802308123       14.1060985019        9.4840154034
  O         4.0758771198       14.0910776698       19.6848128759
  O         7.3781064031        2.8295019896        9.3909574773
  O         7.6628207020        2.7872793291       19.6231218751
  O         7.3686832044        6.6324227358        9.4845319297
  O         7.3127576975        6.5670246964       19.6418908118
  O         7.6512528411       10.3523257589        9.4874769075
  O         7.8256626490       10.3192172788       19.6885123477
  O         7.8288570482       14.1077345453        9.4336171023
  O         7.7643342934       14.0692492105       19.6488186494
  O        11.1021390961        2.8372697854        9.4732389701
  O        11.0756176507        2.7910347174       19.6504400070
  O        11.4673951082        6.5825375871        9.4612822846
  O        11.5154069419        6.5513163465       19.6807877548
  O        11.5896097189       10.3531427697        9.4389267477
  O        11.5625911202       10.3050316987       19.6476935101
  O        11.2115830174       14.1511816635        9.4124177249
  O        11.4897468054       14.1047127765       19.6213503834
  O         1.9614642303        0.9112422076        4.3589691033
  O         2.0583469239        0.9260355534       14.5744588035
  O         2.1171507848        4.6867644277        4.3116423572
  O         2.2069284169        4.7162402547       14.5251470198
  O         1.7763225622        8.4885508888        4.2985514128
  O         1.8243166417        8.5125983479       14.5019385122
  O         1.7198074107       12.2692455597        4.3587443367
  O         2.0539369688       12.2504583122       14.5301958908
  O         5.8643644984        0.9105753275        4.3189518195
  O         5.9743794412        0.9785570114       14.4792728111
  O         5.5203160131        4.7122507946        4.2986798826
  O         5.4794681250        4.6918805469       14.4663591118
  O         5.4660110669        8.4961937608        4.3590362443
  O         5.8079483820       12.2391277396        4.3633165318
  O         9.2718660294        0.9360130984        4.2932103220
  O         9.2561294271        0.9427390090       14.5030036346
  O         9.2090334354        4.7230576019        4.3568309252
  O         9.2887428500        4.7482862558       14.5689233915
  O         9.5496161877        8.4655787066        4.3583869123
  O         9.3636141543        8.5170923067       14.5420130997
  O         9.7167654362       12.2359266140        4.3158399612
  O         9.7232058377       12.2472211924       14.5238623568
  O        12.9527206131        0.9411923475        4.3555941021
  O        13.0335437170        0.9500167173       14.5625100685
  O        13.3069610794        4.6938755115        4.3581284003
  O        13.3055255218        4.7258163708       14.5586923126
  O        13.4603063142        8.4641151104        4.3194203126
  O        13.5657530404        8.4853110445       14.5478680125
  O        13.1032510000       12.2600188022        4.3007815004
  O        13.1025482216       12.2761043908       14.5281475102
  O         0.3357565771       12.2484050849       10.9778758830
  O         1.9718724003        8.3542439861       18.6171564662
  O        13.1382837015       10.3822487651       16.0811568754
  O         3.7847929249        6.7158744189        3.3009898006
  F         5.7969945041       12.3426277000       14.6465267733
  F         5.6637600389        8.4207634569       14.6233709491
  F         3.8119959683       10.4032215442       13.5222561275
  F         7.6813287364       10.3797679233       13.5249406649
    &END COORD
    &KIND O
     ELEMENT O
     MASS    15.99940
    &END KIND
    &KIND Ti
     ELEMENT Ti
     MASS    47.867
    &END KIND
    &KIND F
     ELEMENT F
     MASS    18.998 
    &END KIND
   &END SUBSYS
STRESS_TENSOR ANALYTICAL
&END FORCE_EVAL
\end{verbatim}
\end{tiny}
\end{spacing}

\end{document}